\documentclass[apj]{emulateapj}
\slugcomment{{\sc Accepted to ApJ:} 2013 December 11} 
\usepackage{amsmath}

\slugcomment{}

\shorttitle{IR extinction through RC giants}
\shortauthors{Gonz\'alez-Fern\'andez et al.}

\begin{document}


\title{Infrared extinction in the Inner Milky Way\\ through the red clump giants}


\author{C. Gonz\'{a}lez-Fern\'{a}ndez}
\affil{Instituto de Astrof\'{\i}sica de Canarias \\ E-38205 La Laguna, Tenerife, Spain}
\affil{Departamento de F\'{\i}sica, Ingenier\'{\i}a de Sistemas y Teor\'{\i}a de la Se\~nal \\ Universidad de Alicante, Apdo. 99, 03080, Alicante, Spain}
\email{carlos.gonzalez@ua.es}
\author{A. Asensio Ramos\altaffilmark{1}}
\author{F. Garz\'{o}n\altaffilmark{1}}
\author{A. Cabrera-Lavers\altaffilmark{1}}
\author{P. L. Hammersley\altaffilmark{2}}
\affil{Instituto de Astrof\'{\i}sica de Canarias \\ E-38205 La Laguna, Tenerife, Spain}


\altaffiltext{1}{Departamento de Astrof\'{\i}sica, Universidad de La Laguna, E-38205 La Laguna, Tenerife, Spain}
\altaffiltext{2}{European Southern Observatory, Karl-Schwarzschild-Strasse 2, Garching, 85748, Germany}


\begin{abstract}
While the shape of the extinction curve on the infrared is considered to be set and the extinction ratios between infrared bands are usually taken to be approximately constant, a recent number of studies point either to a spatially variable behavior on the exponent of the power law or to a different extinction law altogether. In this paper, we propose a method to analyze the overall behavior of the interstellar extinction by means of the red-clump population, and we apply it to those areas of the Milky Way where the presence of interstellar matter is heavily felt: areas located in $5^{\circ}<l<30^\circ$ and $b=0^\circ$. We show that the extinction ratios traditionally used for the near infrared could be inappropriate for the inner Galaxy and we analyze the behavior of the extinction law from $1\mu m$ to 8$\mu m$.
\end{abstract}





\section{Introduction}

The extinction law is a useful analytical formula that expresses the amount of light from a given source, in magnitude units, absorbed by the interstellar medium as a function of wavelength. In the infrared, this function $A_{\lambda}$ is usually modeled as a simple power law of the form\footnote{See Section \ref{elaw} for an explanation of the choice of exponents} $A_{\lambda}\propto\lambda^{-\beta}$ \citep[see for example][]{RL85,C89}, although other more complex relations have also been proposed \citep{FM09}.
  
When the bandpasses under consideration are wide (as in most photometric systems, often several hundred angstroms), $A_{\lambda}$ is often computed at a given wavelength representative of each filter. The calculation of this value is problematic: it should be the isophotal wavelength \citep[as defined, for example, in ][]{gol74}, a value that depends on the considered spectral energy distribution. This implies that for every spectral type there is a different isophotal wavelength, and normally an average over several types or the value for a given one (normally a solar-like star) are used instead, or simply the central bandpass' wavelength, much more easily calculated but less representative of the behavior of the extinction under said filter.
  
Ideally, the extinction law should be tailored for every photometric system, but usually a standard relation \citep[such as the one proposed by][]{RL85} is used instead. This relation is normally parameterized through several $A_\lambda/A_\mathrm{V}$ ratios, relating the extinction for a given band to that of the visible $V$ filter. In this paper, we propose a method for the calculation of these coefficients, assuming a $(\lambda,A_{\lambda})$ relation, through the use of red clump giants (RCGs hereafter), and we apply it to 2MASS data \citep{SK06} in the inner Galactic plane ($5^{\circ}<l<30^{\circ}, |b|\leq0.25^{\circ}$), where extinction is more severe. Although 2MASS has coverage of the whole disk, this range complements that of \citet{SH09}, avoiding the innermost Galaxy where source density severely affects 2MASS' depth. Further on, these results will be extended to the mid-infrared using GLIMPSE \citep{B03} data.


\section{The data}

We retrieve all the available data for such fields from the 2MASS, UKIDSS (DR7), GLIMPSE and GLIMPSE-II point source catalogs using the GATOR engine\footnote{http://irsa.ipac.caltech.edu/applications/Gator/}. It should be noted that GLIMPSE data are only available for fields with $|l|\geq10^\circ$; beyond this range, we use GLIMPSE-II.

In these fields, the limiting magnitudes are dictated mostly by source confusion. As the source density is variable, they must be calculated for each pointing. This high stellar density allows us to calculate completeness limits quite easily: using only sources bright enough to be far away from the instrumental limits, we fit the number of stars per magnitude bin with a third degree polynomial. Then, we see at which point prediction and observations differ more than a given threshold. Assuming an 80\% completeness, the typical limiting magnitudes for 2MASS are $m_\mathrm{J}\sim15.5$, $m_\mathrm{H}\sim14.1$ and $m_\mathrm{K_\mathrm{S}}\sim13.3$. For the four bandpasses of GLIMPSE, these values are $[3.6]\sim14.0$, $[4.5]\sim13.7$, $[5.8]\sim12.1$ and $[8.0]\sim11.3$.

Because UKIDSS uses a detector with higher spatial resolution, it is less affected by crowding; thus using data from its Galactic Plane Survey \citep{L08}, we can probe deeper into the Galaxy. Using the procedure outlined before, we see that for our lines of sight, UKIDSS-GPS is complete down to $m_\mathrm{J}\sim18.5$, $m_\mathrm{H}\sim17.5$ and $m_\mathrm{K_\mathrm{S}}\sim16.5$.
  
All along this study, these surveys will be broken into square $0.25~\mathrm{deg^2}$ fields, and we will label them with the coordinates of their centers. Thus, when we refer, for example, to the $(27,0)$ field, we are talking about stars that verify $26.75^\circ\leq l\leq27.25^\circ$ and $|b|\leq0.25^\circ$.

Prior to our analysis, we need to crossmatch both catalogs. To do this, we opt for a critical radius of $1.5"$, which is more than enough to account for the possible differences in astrometry between the two without introducing too many spurious matches.

\section{Calculation of $A_{\mathrm{\lambda}}/A_{\mathrm{V}}$}
\subsection{Isolating the Red Clump Giants}
\label{isored}
RCGs are the dominant population among the giants of our Galaxy, and they have a narrow luminosity function that makes them an ideal subject to study the properties of the Milky Way \citep[see][for a detailed description of the properties of these stars]{CL08}.
    
\begin{figure}[!h]
	\centering
	\includegraphics[width=6.cm,angle=90]{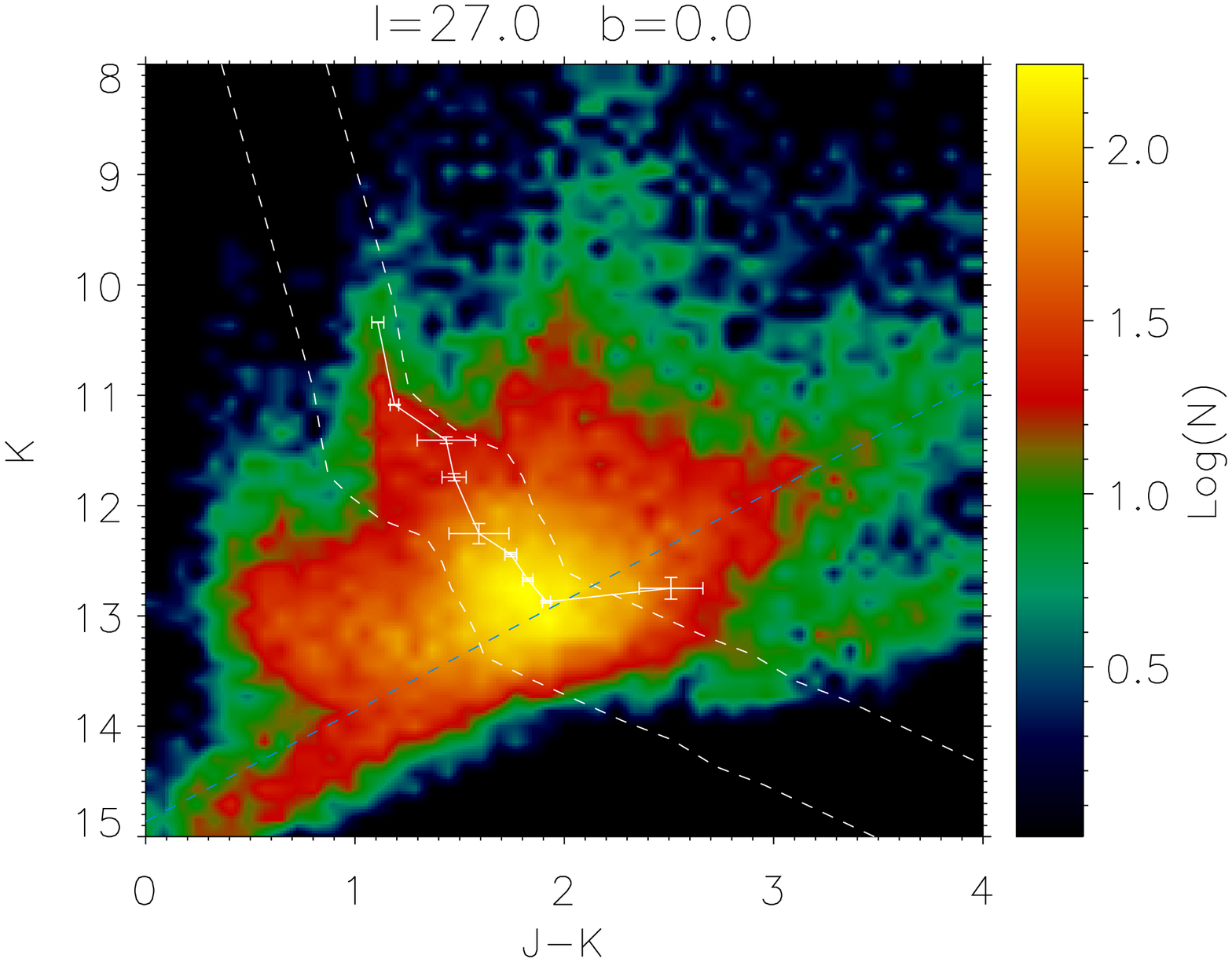}
	\includegraphics[width=6.cm,angle=90]{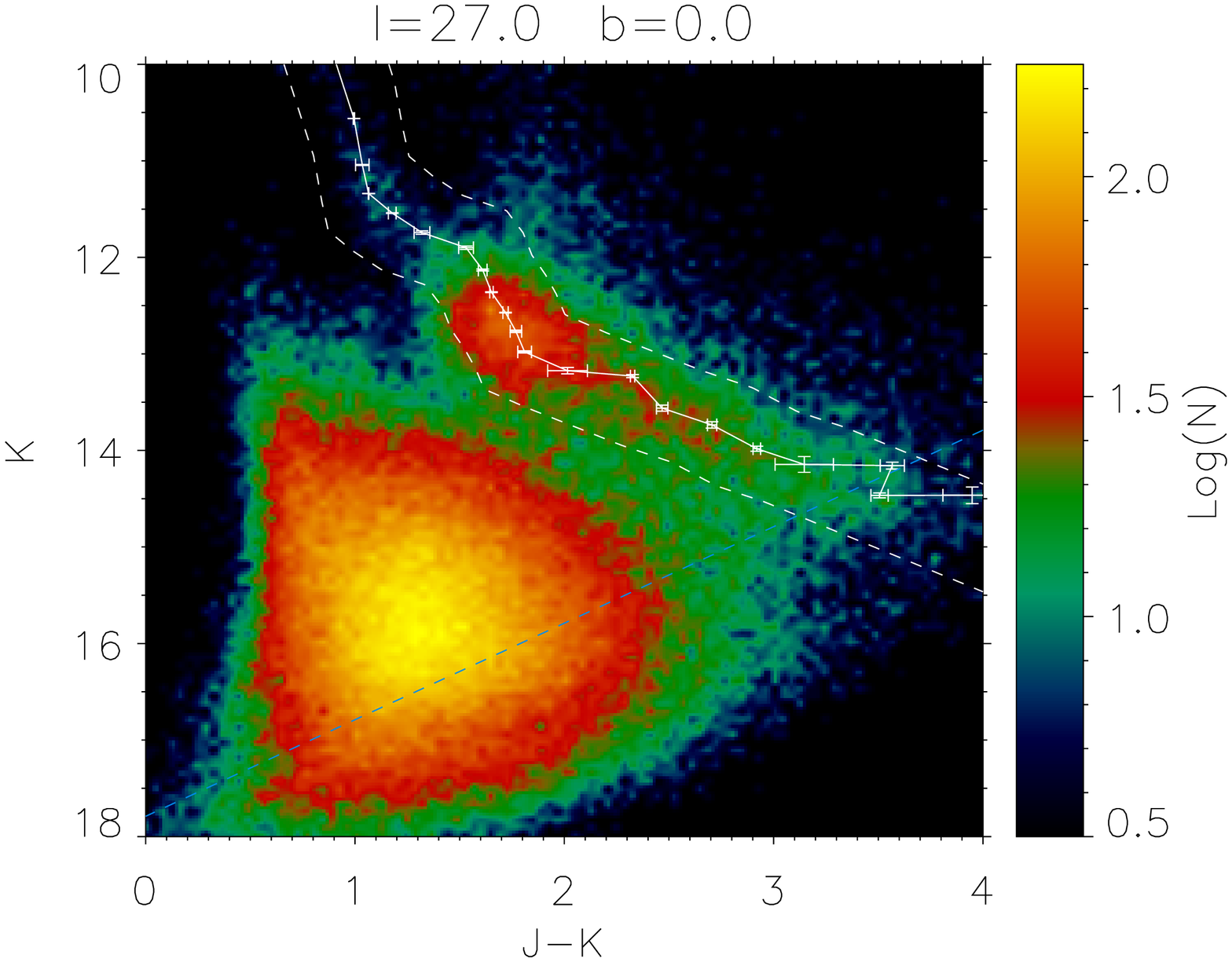}
	\caption{CMD corresponding to the field l=27$^\circ$, b=0$^\circ$ for 2MASS (top) and UKIDSS (bottom). The blue dashed line denotes the completeness limit. The fiducial line tracing the position of an ideal red clump star is marked with a solid white line, while the dashed ones denote the area occupied where these stars are the dominant population. Note that the color scale changes slightly between plots.}
	\label{DCM}
\end{figure}

Due to their abundance, they appear as a clear strip on a color-magnitude diagram (CMD henceforth, see Figure \ref{DCM}), particularly in disk fields. This makes it possible to isolate them from other populations. The method we use to do this is an evolution of that in \citet{CL05}. To determine the position of a RCG in the  ($J-K,m_{\mathrm{K}}$) space, we first use the extinction provided by the SKY model \citep{wainssky} and the extinction law of \citet{RL85}. With these, a theoretical trace is obtained for the RCGs. The CMD is then sliced into strips across the magnitude axis, and using this trace as a reference, we rotate the CMD so that for each strip we generate a new coordinate system in which the $Y$ axis is tangent to the RCG trace. This allows us to map precisely the position of the RCGs even in regions of high extinction, where they would occupy an almost horizontal strip in the ($J-K,m_{\mathrm{K}}$) diagram. We then find the maximum of the distribution of stars along the $X$ axis, and transform this point back to the ($J-K,m_{\mathrm{K}}$) system, defining the fiducial position of the red clump at the magnitude under consideration.
    
Once the position $(J-K,m_{\mathrm{K}})$ of the red clump is measured over the entire CMD (the solid white line in Figure \ref{DCM}), we are ready to filter the RCGs. We compute several pairs of traces parallel to ($J-K,m_{\mathrm{K}}$)$_{RC}$ but displaced a given distance $r$ above and below it (the white dashed lines in Figure \ref{DCM}). We then calculate the derivate $dN/dr$ of the number of stars contained between each pair. The minimum of this derivate yields the region that optimizes the number of giants: as we move away from ($J-K,m_{\mathrm{K}}$)$_{RC}$ the number of giants decays until we start introducing dwarf disk stars, at which point $dN/dr$ begins to rise again.

Within these frontiers fall most of the RCGs along the line of sight, and so in this range they are the most abundant population. Dwarfs are the major contaminant for our analysis (this will be explored in Section \ref{elaw}), but the lower limit of the RCG trace will filter them out, as their fainter absolute magnitudes puts them below this region; in fact, depending on the field, $60-80\%$ of all the stars will be discarded with this filter.

\begin{figure}[!h]
	\centering
	\includegraphics[width=6.cm,angle=90]{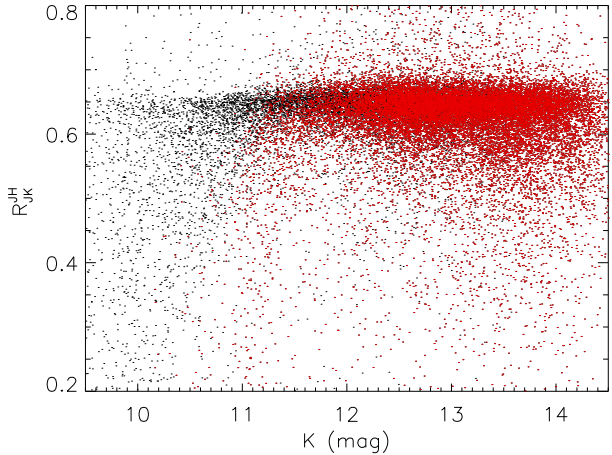}
	\caption{Sample of $E(J-H)/E(J-K)$ for selected sources in Figure \ref{DCM}. The RCGs (i.e., those stars falling in the area delimited by the striped lines in Figure \ref{DCM}) are marked with red dots, while all the late giants (i.e., those stars beyond the lower limit) are plotted with black dots. Compare with Figure \ref{sintejhjk}.}
	\label{RGCsample}
\end{figure}

The upper boundary, on the other hand, cuts out sources that at the same magnitude as the RCGs appear redder. These will be either RCGs that are under particularly high extinction, causing them to appear to the lower right of the main body of RCGs on a CMD, or giants (as they still need to be intrinsically bright) with later spectral types. Since we don't want our analysis to be biased against sources under high extinction, we will only use the lower boundary as a filter. This has the disadvantage of introducing late-type giants in our sample, but as can be seen in Figure \ref{sintejhjk}, if we consider them to be RCGs, all late giants have very similar color excess ratios, so their effect will be to increase the dispersion of our sample without changing its mean value, as can be seen in Figure \ref{RGCsample}.

There will be some contaminants passing our filtering. In the dimmer parts of the CMD the interlopers will be dwarfs, as it is discussed in \citet{lcgiants} and \citet{CL05} (Figure 6 of this paper is particularly relevant to this discussion). The fraction of these contaminants is only significant close to the completeness limit (in this case, imposed by the J band, as can be seen in Figure \ref{DCM}). In heavily obscured fields this effect becomes lower, as the reddening shifts the giants farther to the right on the CMD while dwarfs are less affected, so the dwarf-giant separation is improved.
    
With this we guarantee that most of the dwarfs are filtered out of our sample while the majority of the late giant population (roughly beyond G8III) will still be present. As can be seen in Figure \ref{RGCsample}, this more ample selection also includes more reddened bright giants that extend the sample to lower magnitudes and potentially to  more heavily obscured environments.

\subsection{Modeling the Red Clump}
\label{rgmodel}
\begin{figure}[!h]
	\centering
	\includegraphics[width=6.cm,angle=90]{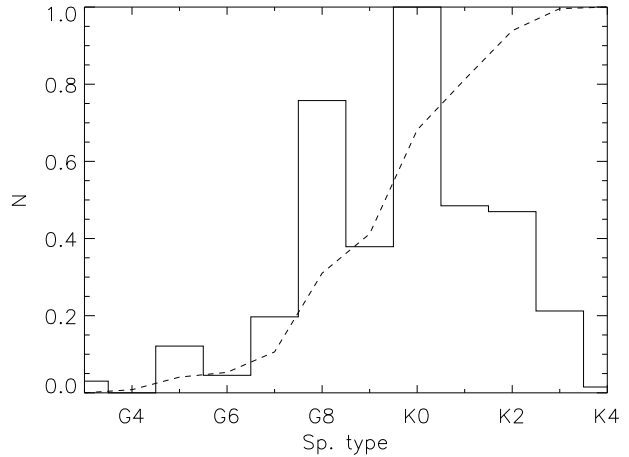}
	\caption{Spectral type histogram and cumulative distribution function of the sample of RCGs from \citet{algc}.}
	\label{alves}
\end{figure}

At the core of our method lies the necessity of a spectral template representative of the stars isolated in the previous step. As shown in \citet{CL08}, the late giant population is dominated by RCG stars. We examine this population using the data of \citet{algc}, consisting of 238 RCGs identified in the Hipparcos catalog. As can be seen in Figure \ref{alves}, the mean and modal spectral type of the sample is K0; this distribution has mean intrinsic colors of $(J-H)_0=0.54\pm 0.09$, $(H-K)_0=0.12\pm 0.09$ and $(J-K)_0=0.65\pm 0.11$, where the errors are simply the standard deviation of the sample for each color. This includes both the intrinsic dispersion in color associated with the population and other astrophysical factors such as the variation introduced by interstellar extinction. The expected intrinsic dispersion in color is closer to $\pm0.05$, as it is shown in \citet{stra09}. We will use this value as the uncertainty for all the intrinsic colors in this work. This will be a slight overestimation in the mid-infrared, where star to star differences tend to be smaller.
	
As we perform synthetic photometry to invert the extinction law, we choose a model from \citet{camodel} representative of a K0III star. According to \citet{stramodel}, the best synthetic spectra for this purpose has solar metallicity, $T_\mathrm{eff}=4750\mathrm{K}$ and $\log(g)=3.0$. We cross-check the magnitudes derived by us with \citet{hephot} and \citet{tokuphot} (with typical differences of a few hundredths of magnitude, associated with the different models chosen for Vega), and obtain for our K0III template the intrinsic colors in all the photometric systems used (Table \ref{k0syn}), consistent, within the errors, with those derived for the near-infrared for the \citet{algc} sample.
	
\begin{table}
		\caption{Synthetic colours obtained for a K0III star.
		\label{k0syn}}
		\begin{center}
		\begin{tabular}{ccc}
			\tableline
			\tableline
			&\multicolumn{2}{c}{Phot. System}\\
			\tableline
			Colour & 2MASS & UKIDSS \\
			\tableline
			J-H&0.58&0.53\\
			H-K&0.07&0.08\\
			J-K&0.65&0.61\\
			$[3]$-K&-0.04&-0.05\\
			$[4]$-K&0.01&0.00\\
			$[5]$-K&-0.01&-0.02\\
			$[8]$-K&-0.06&-0.07\\
			\tableline
		\end{tabular}
		\end{center}
	\end{table}

\subsection{The Extinction Law}
\label{elaw}
   
In this work we use the law described in \citet{FM09}. The authors find that the extinction along several lines of sight can be modeled by:
\begin{equation}
\label{massa}
	k(\lambda-V)=\frac{E(\lambda-V)}{E(B-V)}=
	\frac{A_{\lambda}-A_{V}}{A_{B}-A_{V}}=k_{IR}\frac{1}{1+(\lambda/\lambda_{0})^{\alpha}}-R_V
\end{equation}
where $\lambda_{0}=0.507~\mu m$ and $k_{IR}$ are constants. It follows that for any given three filters at wavelengths $\lambda_1$, $\lambda_2$ and $\lambda_3$, the ratio of color excesses fulfills:
\begin{equation}
\label{rat}
	R^{12}_{23}=\frac{A_1-A_2}{A_1-A_3}=\frac{m_1-m_2-(M_1-M_2)}{m_1-m_3-(M_1-M_3)}=f(\alpha)
\end{equation}
where $(M_1-M_2)$ is the intrinsic color represented as the difference of two absolute magnitudes.
    
Although this expression is model-dependent, in most of the extinction laws found in the literature $A_{\lambda}$ is proportional to a power of $\lambda$, and so the method is applicable to any of them. Along these lines, we will refer to the exponent of Equation (\ref{massa}) as $\alpha$, while $\beta$ will always be used for the traditional $\lambda^{-\beta}$ model.
    
\begin{figure}[!h]
	\centering
	\includegraphics[width=6.cm,angle=90]{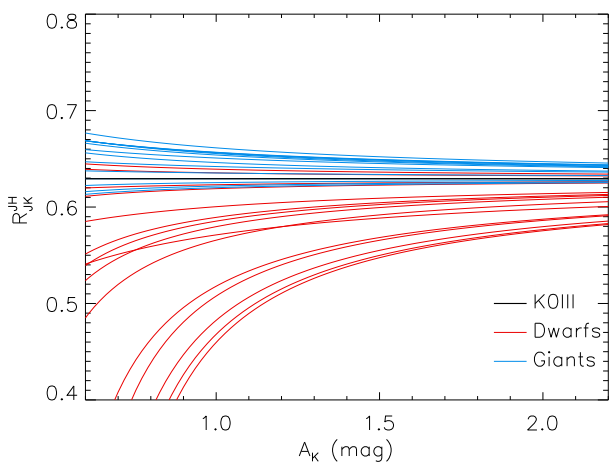}
	\caption{$R^\mathrm{JH}_\mathrm{JK}$ as a function of $A_\mathrm{K}$ for several spectral types of giants (blue lines) and dwarfs (red lines). The color excess ratio is computed against the intrinsic colors of a K0III star (black line).}
	\label{sintejhjk}
\end{figure}

To obtain $R^\mathrm{JH}_\mathrm{JK}$ we have to assume an intrinsic color for the population under analysis. As we have shown in Section \ref{isored}, while our sample is  dominated by RCGs, there will be interlopers, particularly at dimmer magnitudes. This it is convenient to analyze the results obtained for different stellar populations when calculating the color excess ratio assuming a fixed intrinsic color, so that we can interpret diagrams similar to that of Figure \ref{RGCsample}. Using the absolute magnitudes for different dwarfs (from B1 to late M types) and giants (G2 to M7) from \citet{wainssky} and the extinction coefficients from \citet{RL85} we can construct the synthetic diagram of Figure \ref{sintejhjk}. In  this plot, each trace represents the effect of increasing the extinction for a given population while assuming a constant distance. The effects of the mixture of populations translate mainly into two features: giants occupy a rather narrow band around the locus of the K0III, more tightly packed below it; very late dwarfs (beyond M2) are indistinguishable from giants, while younger ones tend to spread over the lower part of the diagram (but these ones are preferentially filtered out in the selection scheme outlined in sect. \ref{isored}). Once we take into account the effect of distance spread, we obtain a diagram similar to that of Figure \ref{RGCsample} in which the area of the plot below the assumed $R^\mathrm{JH}_\mathrm{JK}$ value for an RCG is more or less uniformly populated by interloping dwarfs, while giants occupy a rather narrow band around this value.
    
Although as has been discussed before, there are some interlopers, the sample will always be dominated by RCGs, down to the completeness limit of 2MASS. If we assume that the population distribution from \cite{algc} can be extrapolated to the inner Milky Way, the modal $R^\mathrm{JH}_\mathrm{JK}$ will be associated with the assumed K0III model. This enables us to invert the exponent $\alpha$ of the subjacent extinction law from color excess ratios.
    
Before proceeding to the analysis of color excess, we need to take into account the variation of the intrinsic color of the RCGs with metallicity. As this value decreases, RCGs become slightly bluer, but the measured color variation is very low, around $0.02$ mag in $(J-K)_0$ for a difference of $0.1$ dex \citep[see][for a discussion on this]{lcgiants}. In \citet{cgf08}, the authors find that the variation of the mean metallicity for a sample of RCGs in these inner disc regions is low, around $0.2~dex$. Furthermore, this effect is already present in \citet{stra09}, and since our assumed uncertainties are the ones derived there, they will include the effect of variable metallicity.

Sadly, there is a lack of similar studies for the mid-infrared, although since our population selection is in the near-infrared, we can guarantee that the sample is the same. Extinction effects are less severe at these wavelengths, and as the spectral energy distribution of the stars starts to follow closely a Raleigh-Jeans law, differences in intrinsic color are also mitigated. In fact, using the measurements from \citet{tokunir} we can see that for a uniform distribution of the expected populations in our sample (roughly, giants with types later than G8) the dispersion in $(J-K)_0$ is 0.22 magnitudes, while for $(K-L)_0$, $(K-L')_0$ and $(K-M)_0$ it drops to $0.05~\mathrm{mag}$, although these bands are between $3.0$ and $5.0~\mathrm{\mu m}$, it is reasonable to expect that this behavior can be extrapolated to the GLIMPSE photometric system. As our sample is heavily dominated by a small range of spectral types, even without deriving synthetic colors in the MIR, it is reasonable to assume that the uncertainties derived for 2MASS colors will also serve as an envelope for the GLIMPSE intrinsic colors.

\section{Results}
\subsection{Spatial Variability of the Extinction Model}
\label{statana}
For each field we construct a diagram as the one in Figure \ref{RGCsample}. We choose the (J-H) versus (J-K) ratio because it minimizes the effect of errors. Standard propagation of uncertainties over Equation \ref{rat} yields:
\begin{equation}
\label{error}
	\frac{\Delta(R^{12}_{13})}{R^{12}_{13}}\propto\\
	\frac{1}{\left[m_1-m_2-(M_1-M_2)\right]\cdot\left[m_1-m_3-(M_1-M_3)\right]}
\end{equation}
Assuming that the errors in magnitude are equal for all three filters and that the uncertainties in the intrinsic colors are similar, $R^\mathrm{JH}_\mathrm{JK}$ puts the greater excess on the denominator of Equation \ref{error}, and thus yields a smaller relative error.
    
\begin{figure}[ht!]
	\centering
	\includegraphics[width=6.cm,angle=90]{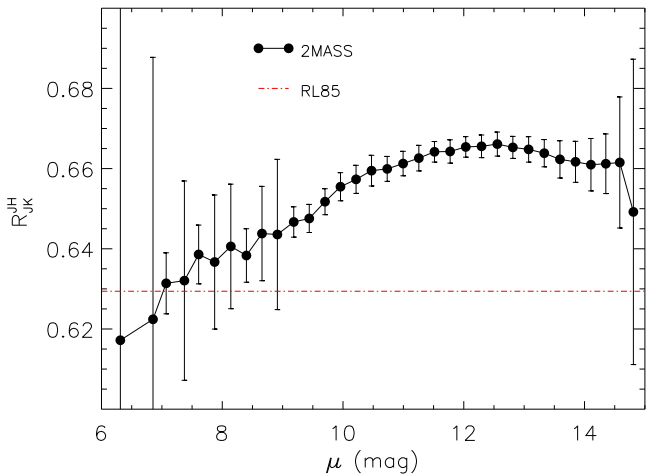}
	\caption{Variation of $R^\mathrm{JH}_\mathrm{JK}$ with distance modulus, averaged over all the lines of sight. Black dots are for 2MASS stars with good photometry (grade A in all three bands), and the red dashed line marks the ratio predicted by \citet{RL85}. \citet{C89} predicts $R=0.54$, outside the plotting range.
	\label{ratall}}
\end{figure}

In order to see if there is evidence of a variable $R^\mathrm{JH}_\mathrm{JK}$ along the line of sight, we use the distance modulus $\mu$ as a proxy for distance, as we do not want to assume an a priori extinction law (this would allow us to transform from magnitude to distance). At each of the Galactic coordinates we slice our sample into bins of 0.25 mag in $\mu$, and we obtain the modal value of this ratio for each bin. We require all the stars included in these bins to have good photometry (i.e. {\it AAA} grade in 2MASS), so to avoid possible saturation effects. We average over all the lines of sight, so that we can see how the color excess ratio behaves with $\mu$ even at the bright end of our data (as these stars will be near us, we need to cover a wider solid angle to get a meaningful sample). The results are plotted in  Figure \ref{ratall}.
      
We can see in Figure \ref{ratall} that it seems that while for stars with large distance modulus, probing the inner Galaxy, the color excess ratio remains more or less constant, while at lower $\mu$ it tends towards the values predicted by \cite{RL85}. Although is tempting to assume that this is the result of a variable extinction law, a model with spatially variable coefficients bears an statistical tax that we need to take into account before choosing what function best fits our data. It is possible to evaluate this with a variation of the Minimum Description Length (MDL) principle \citep{riss98, AR06}, based on a clustering algorithm.
      
As we don't know the dependency $R^\mathrm{JH}_\mathrm{JK}$ with $\mu$, or even if there is one, we cannot construct a family of models to check against the observations. We rely instead on substituting this family of models by successive clusterings of the data using Ward's method. At each iteration we partition the dataset in $n$ clusters (with $n$ increasing from 1 to 10), and our "model" consists in substituting each cluster by its mean. We can calculate the total length of the message needed to transmit this model and the significant residuals (i.e. those that are over their respective dispersions), and see how this length varies with the number of clusters.
      
If all the differences in the dataset fall below the dispersion of the data, the optimal description of the system will consist of a single cluster, but if there is significant structure above the error level, even if $R^\mathrm{JH}_\mathrm{JK}$ depends on some unknown variables, it would show up through an optimal partition consisting of several clusters, in a way that minimizes the more significant residuals.
      
\begin{figure}[ht!]
	\centering
	\includegraphics[width=6.cm,angle=90]{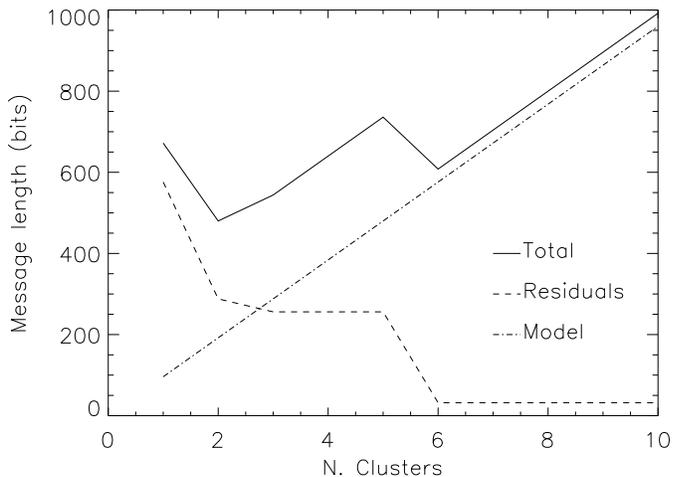}
	\protect\caption[ ]{Description length of the $R^\mathrm{JH}_\mathrm{JK}$ data for Figure \ref{ratall}. The solid line is the total description length for model and residuals, plotted separately with dashed and dot-dashed lines respectively.
	\label{mdl}}
\end{figure}

The results of this procedure are shown in Figure \ref{mdl}, in which the message length for each successive division of the datasets is plotted. The optimal partition requires two clusters: the first containing values close to \citet{RL85} and the second including the higher $R^\mathrm{JH}_\mathrm{JK}$ present at dimmer magnitudes. These calculations also show that the regime change occurs between $\mu\sim9$ and $\mu\sim10$.

In view of this result, it is tempting to conclude that the extinction law from \citet{RL85} is applicable to the first kiloparsecs along the line of sight and that for the innermost Galaxy we need a new expression. Although the authors of this study sampled high extinction stars in order to make their results as universal as possible, only two of their stars have measurements in all three infrared filters used here, and they have values of $R^\mathrm{JH}_\mathrm{JK}$ (0.63 and 0.69) that span a range compatible with our results. Furthermore, \citet{RL85} assumed a priori that the extinction law is constant in order to be able to derive color excesses and $A_\mathrm{V}$ for stars without photometry in the visible, which biased their calculation against stars with non-standard behavior. The data at hand show clearly that the extinction law indeed varies along the line of sight, yet to properly characterize this (and to find where this change in behavior occurs) more data are needed, as it would be necessary to disentangle extinction and distance effects.
		      
It should be noted that using the color excess ratios for all the lines of sight introduces some extra dispersion. One should expect the amount of interstellar material along the line of sight to have a dependence with galactocentric distance, but we are averaging over distance modulus. This will result in some mixing of stars at various galactocentric distances and under different amounts of obscuration but the same apparent magnitude. Yet this does not affect our conclusions. First, our range in Galactic latitude is modest and so for a given $\mu$ all the stars will come from a rather small range in $R_{GC}$. Second, the final result of this effect should be an increase in the dispersion at each $\mu$, and so it could mask any underlying spatial variation, but we are able to detect it even above this artificially increased variance. In fact, if we compare the dispersions of the averaged values to those at each line of sight we find that the former are of the same order of magnitude, if not slightly smaller, than the latter. This indicates us that the possible extra variance introduced remains below the one intrinsic to the use of 0.25 bins in $\mu$.
      
There are few stars with $\mu<9$, roughly $5\%$ of the whole sample (hence the need to average over all the lines of sight in Figure \ref{ratall}). Being so, we can calculate a mean value for $R^\mathrm{JH}_\mathrm{JK}$ (or any other excess ratio) for each line of sight, excluding these stars. With these ratios, we can look for differences between lines of sight. With the same analysis used before, {we find that the optimal partitioning of the sample only requires one cluster, implying that} there is no evidence of a large scale variation with Galactic latitude in the range $5^\circ<l<30^\circ$.

\begin{figure}[ht!]
	\centering
	\includegraphics[width=6.cm,angle=90]{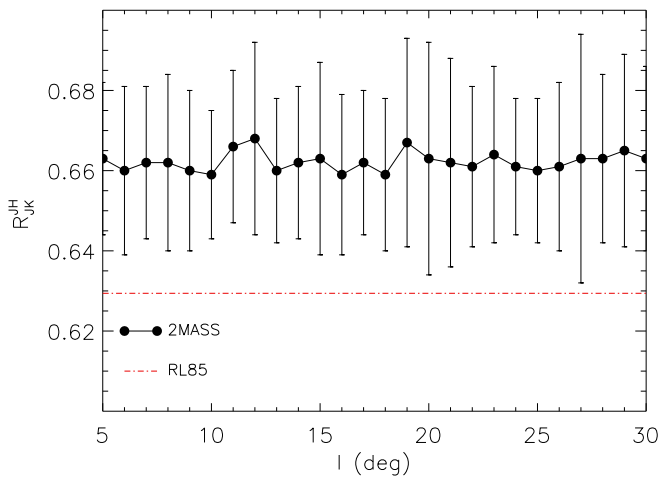}
	\protect\caption[ ]{Variation of $R^\mathrm{JH}_\mathrm{JK}$ with Galactic longitude. As can be seen, any azimuthal gradient is swallowed by the intrinsic dispersion within a given line of sight.
	\label{ratvsl}}
\end{figure}

Two facts about this result should be highlighted. First, our method is only sensible to large scale variations. Other studies, such as \citet{G09}, find significant changes on the slope of the extinction law. These changes occur at very small scales, around $5"$. As our analysis deals with fields $0.25$ degrees across, such variations will be blurred. Second, the key of our calculations is the derivation of a meaningful dispersion measurement for the color excess ratios. Looking at Figure \ref{ratvsl} one could conclude that a $\sigma\sim0.02$ is too high a value and that this will mask any significant variation between lines of sight. Indeed, as at each Galactic longitude we measure $R^\mathrm{JH}_\mathrm{JK}$ using a large number of stars, the error will surely be well below the 0.06 expected from simple error propagation. But in constructing Figure \ref{ratvsl} we assume that $R^\mathrm{JH}_\mathrm{JK}$ is constant at each line of sight, and so the dispersion observable between all the stars for a given field is a good measurement of the precision of our method.

It is also worth noting that we limit ourselves to a rather narrow range in Galactic longitude. As we move away from the Galactic center, the regions of the Galaxy we sample become less homogeneous as, for example, some of them will be almost tangent to spiral arms while others will mostly cross only inter-arm regions. This results in a more much divergent behavior of the extinction law, as can be seen, for example, in \citet{Z09} or \citet{SH09}.

If there is no evidence of a variation with $l$, we can derive a mean $R^\mathrm{JH}_\mathrm{JK}$ for each line of sight. This value will be dominated by  stars with dimmer magnitudes (hence farther away from us), and we can use it to check whether there is an azimuthal variation. As can be seen in Figure \ref{ratvsl}, any variation with Galactic longitude is too small to be detected with this our method. Although for a given line of sight the dispersion is relatively high, around $0.02$, the modal $R^\mathrm{JH}_\mathrm{JK}$ remains notably homogeneous for all the lines of sight.

Consequently, it is meaningful to derive a mean value of the color excess ratios for all the lines of sight (table \ref{rats}), including all the stars. The uncertainties represent only the dispersion between lines of sight. On top of it there is a systematic $\pm0.07$ associated with the error on the intrinsic colors for the RCGs.

\begin{table}
		\caption{Mean color excess ratios derived from our sample.
		\label{rats}}
		\begin{center}
		\begin{tabular}{ccc}
			\tableline
			\tableline
			&\multicolumn{2}{c}{Phot. System}\\
			\tableline
			Ratio & 2MASS & UKIDSS \\
			\tableline
			(J-H)/(J-K)&0.661$\pm$0.003&0.644$\pm$0.013\\
			(H-K)/(J-H)&0.517$\pm$0.006&0.56$\pm$0.03\\
			(H-K)/(J-K)&0.343$\pm$0.003&0.360$\pm$0.013\\
			\tableline
		\end{tabular}
		\end{center}
	\end{table}

\subsection{Derivation of the exponent}
\label{devexp}
    
The total extinction $A_\lambda$ for a given filter with transmission curve $T(\lambda)$, in the range $(\lambda_0,\lambda_1)$ is given by:
\begin{equation}
\label{refext}
	A_{\lambda}=-2.5\cdot \log \frac{\int^{\lambda_{1}}_{\lambda_{0}} T(\lambda) f(\lambda) 10^{-0.4\cdot
	A(\lambda)}d\lambda}{\int^{\lambda_{1}}_{\lambda_{0}}T(\lambda)f(\lambda) d\lambda}
\end{equation}   
where $f(\lambda)$ is the spectral energy distribution of the source under consideration and $A(\lambda)$ the extinction model assumed. The derivation of $A(\lambda)$ from $k(\lambda-V)$ (Equation \ref{massa}) requires knowledge of the proportionality constant $k_{IR}$; according to \citet{FM09} it can be parameterized as:
\begin{equation}
\label{kir}
	k_{IR}=0.349+2.087\cdot R_{V}\\
\end{equation}
But then a value for the selective to total extinction ratio $R_V$ and an absolute value for the extinction at some wavelength, given for example through $A_\mathrm{V}$, are needed to evaluate $A(\lambda)$. One can carry out Bayesian inference over the three parameters, $\alpha$, $R_\mathrm{V}$ and $A_\mathrm{V}$, using all the color excess ratios and their $\sigma$ from Table \ref{rats} and an uninformative prior for all three. Assuming Gaussian errors for the observations, we can set up a likelihood function for the values of $(A_\mathrm{J}-A_\mathrm{H})/(A_\mathrm{J}-A_\mathrm{K})$.

\begin{figure}[ht!]
	\centering
	\includegraphics[width=6.cm,angle=90]{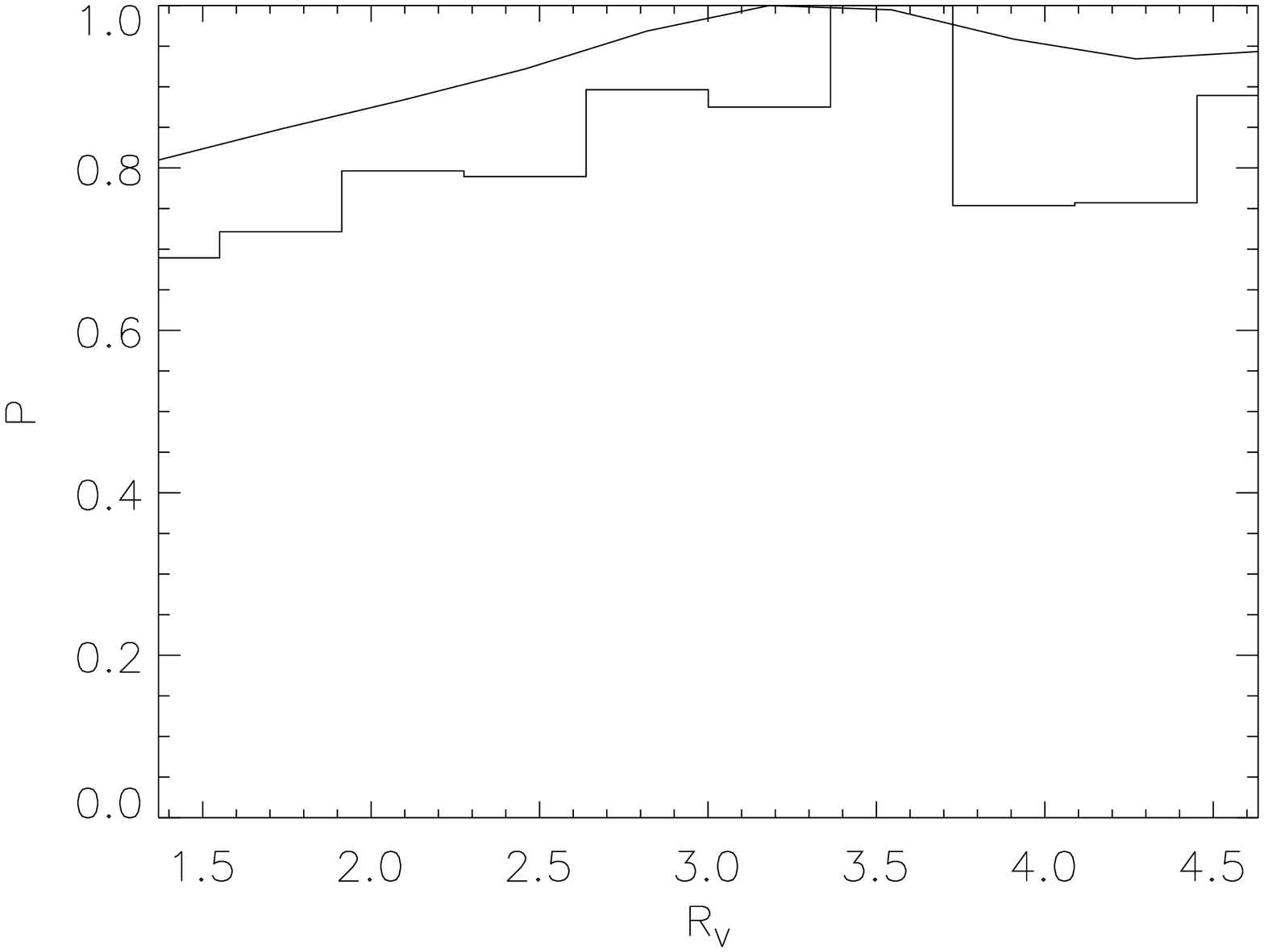}
	\includegraphics[width=6.cm,angle=90]{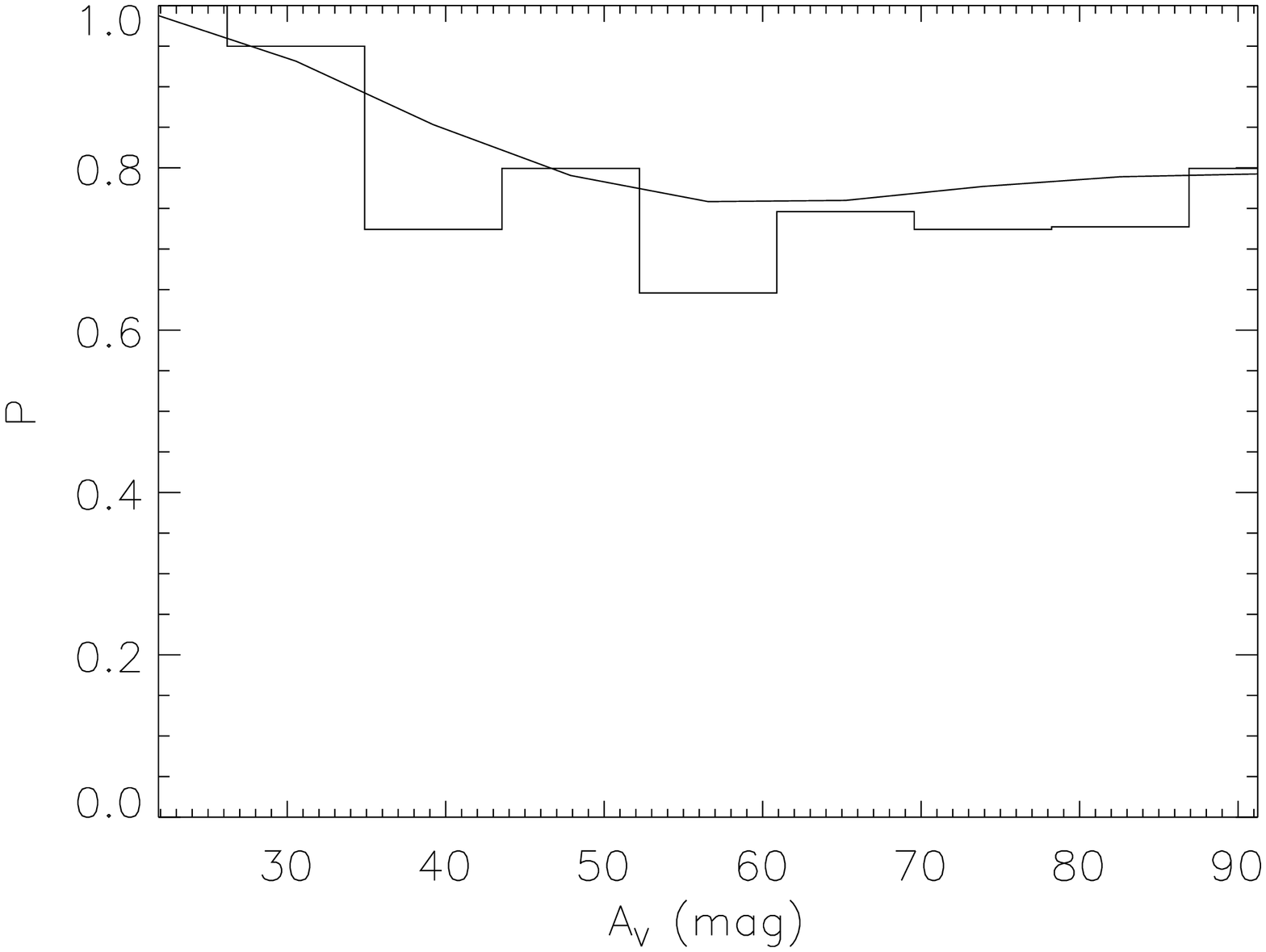}
	\includegraphics[width=6.cm,angle=90]{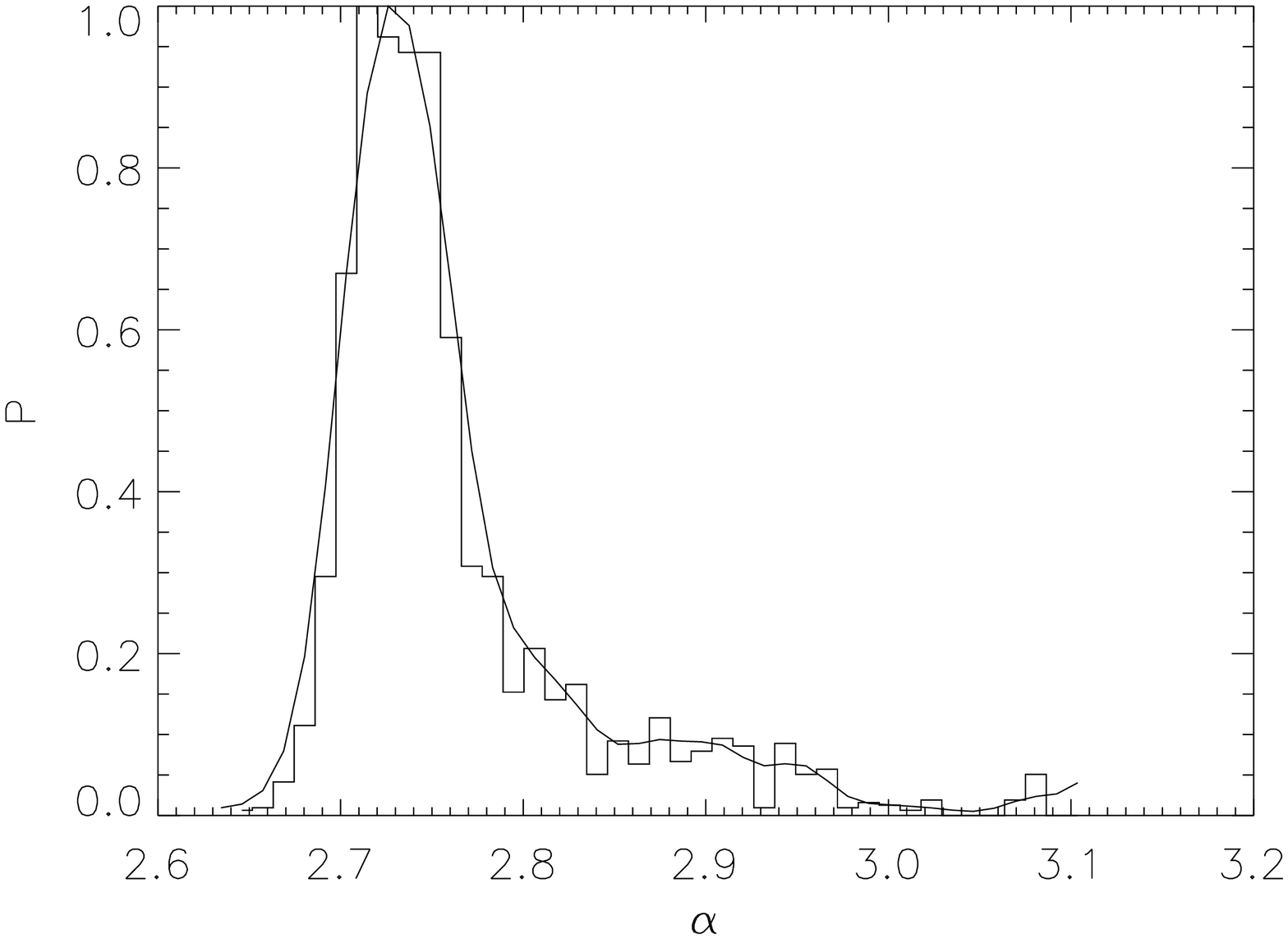}
	\protect\caption[ ]{Marginalized posterior distribution, normalized to its maximum, for $R_{V}$ (top), $E(B-V)$ (middle) and $\alpha$ (bottom). The behavior of the first two indicates that both parameters have little influence in the determination of the optimal exponent, even in a wide range of variation, enough to cover the values expected through the whole Galaxy.
	\label{posterior}}
\end{figure}

Using the selected SED in \ref{rgmodel} and the total efficiency of the photometric system in each passband $T(\lambda)$, a value for $A_{\lambda}$ can be obtained through numerical integration. To evaluate a given excess ratio $R^{12}_{23}$ this operation needs to be performed three times. To speed up the calculation, the sampling of the full posterior distribution is carried out using a Markov Chain Monte Carlo method, which automatically computes the marginal posterior distributions\footnote{For a brief introduction to Bayesian inference, see \citet{LO99}} from which statistically relevant confidence margins can be correctly defined. Uniform priors are set over the three parameters in the ranges $1<R_V<5$, $10<A_V<100$ and $2<\alpha<5$ (the same range was used for $\beta$) and a Markov Chain of length 50000 is computed. Using the ratios from Table \ref{rats}, the marginal posteriors of Figure \ref{posterior} are obtained. As the marginal posterior distributions for $R_V$ and $A_V$ are compatible with the prior distributions, these parameters are not constrained at all using $R^\mathrm{JH}_\mathrm{JK}$, just as expected. On the other hand, $\alpha$ is very well constrained by the data:
\begin{eqnarray*}
	\alpha_{NIR}= 2.76\pm0.04\\
\end{eqnarray*}
      
It should be noted that this uncertainty derives only from the integration of the posterior distribution taking into account the variance derived from averaging all our lines of sight, as it is shown in Table \ref{rats}. This is not the only source of error, as these will be dominated by the systematics introduced by the uncertainties over the intrinsic colors of the RCGs. As it is discussed in Section \ref{rgmodel} these are typically of $0.05$ mag. This introduces an error of $0.07$ on any color-to-color ratio. If we feed this into the Bayesian inversion, we come up with a systematic of $\pm 0.15$ to be added to the uncertainty of all the derived exponents.

\subsection{Extension to the MIR}
\label{exmir}

It is possible to extend this analysis to the mid-infrared, using data from GLIMPSE. The limiting magnitudes in the two reddest MIR bands are noticeably smaller than those of 2MASS. This implies that, for our lines of sight, while around $99\%$ of the RCGs detected in the NIR have a counterpart at $3.6$ and $4.5~\mu m$, this fraction drops to $86\%$ at $5.8~\mu m$ and $60\%$ for the $8.0~\mu m$ band. This implies that our probe into the Milky Way will be shallower at these wavelengths, although this effect is mitigated partly due to the transparency of the interstellar material at these magnitudes. Being so, the limiting magnitude is mostly dominated by crowding effects, and in fact the ratio of detected RCGs at $8.0~\mu m$ drops from $77\%$ at $l=30^\circ$ to $40\%$ at $l=5^\circ$. But as we will use azimuthally aggregated values for the color ratios, these differences are smeared out.
      
We select our color ratios following \citet{I05}, and we use the synthetic intrinsic colors from table \ref{k0syn}. {There are two caveats for this analysis. First, we lack an observational confirmation of the suitability of the stellar model used redward of 3$\mu m$, although, as we have seen, the variations of intrinsic color in this regime are very low. Second, the law from \citet{FM09} has been derived for shorter wavelengths, and the validity of an extrapolation is by no means guaranteed. Being so, results derived here values should be interpreted with care.}
      
Averaging for all the lines of sight, we obtain the values in Table \ref{ratcolmir}. Using these and the same Bayesian inversion described previously, we obtain:
\begin{eqnarray*}
	\alpha_{MIR}= 2.73\pm0.04\\
\end{eqnarray*}

	\begin{table}[!h]
		\caption{Colour excess ratios for the crossmatch of GLIMPSE data with 2MASS.
		\label{ratcolmir}}
		\begin{center}
		\begin{tabular}{cc}
			\tableline
			\tableline
			Ratio & Value\\
			\tableline
			([3]-K)/(J-K)&-0.289$\pm$0.012\\
			([4]-K)/(J-K)&-0.290$\pm$0.010\\
			([5]-K)/(J-K)&-0.277$\pm$0.014\\
			([8]-K)/(J-K)&-0.310$\pm$0.010\\
			\tableline
		\end{tabular}
		\end{center}
	\end{table}
		 
\subsection{$\lambda^{-\beta}$ law}
\label{exbet}
    
The same calculations described before can be extended directly to a law following $A_\lambda\propto\lambda^{-\beta}$. We can fully parametrize this relation following \citet{FM09}:
\begin{equation}
\label{beteq}
	k(\lambda-V)=\frac{A_\lambda-A_V}{A_B-A_V}=k_{IR}\lambda^{-\beta}-R_{V}
\end{equation}
And so again $A(\lambda)$ has a dependence on $A_\mathrm{V}$ and $R_\mathrm{V}$. Even if they share notation, the proportionality constants $k_\mathrm{IR}$ are different for this and the aforementioned law, and so a new calibration of the ($k_\mathrm{IR}$,$R_\mathrm{V}$) is needed. Using the data from \citet{FM09} we reach:
\begin{equation}
	k_{IR}=-1.045+0.721\cdot R_{V}
\end{equation}
After this calculation, we can invert the values for $\beta$ from the previously obtained color excesses:
\begin{eqnarray*}
	\beta_{NIR}= 2.52\pm0.06\\
	\beta_{MIR}= 2.64\pm0.07\\
\end{eqnarray*}

\section{Discussion}
\subsection{Comparison between Extinction Laws}
    
To test how both laws describe the overall behavior of the interstellar material, we have to repeat the previous analysis for the whole wavelength range. Repeating the Bayesian inference scheme used in previous sections with all the values from Tables \ref{rats} and \ref{ratcolmir}, we obtain a posterior distribution similar to that of Figure \ref{posterior}; a Gaussian fit to the posterior distribution gives us the values for all our wavelength range:
\begin{eqnarray*}
	\alpha_{total}=2.73\pm0.02\\
	\beta_{total}=2.52\pm0.02\\
\end{eqnarray*}
As discussed previously, there is a systematic $\pm0.15$ to be added to these errors.
		 
\begin{figure}[ht!]
	\centering
	\includegraphics[width=6.cm,angle=90]{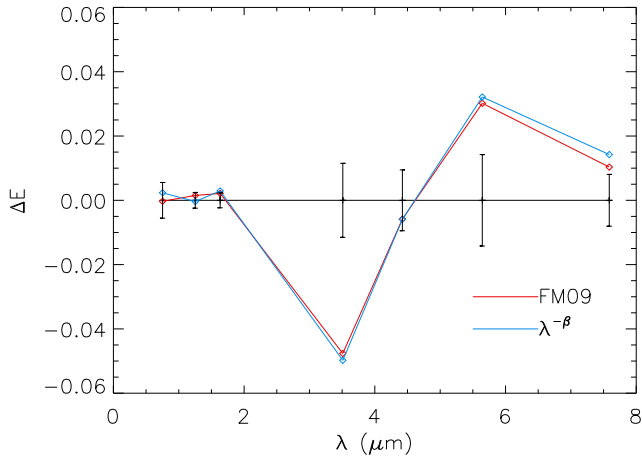}
	\protect\caption[ ]{Differences in the color excess ratios $\Delta E$ between the modeled values and the measured ones for the 2MASS photometric system. The first two points (at arbitrary wavelengths $\lambda\sim0.5\mu m$ and  $\lambda\sim1.5\mu m$) represent $\frac{E(H-K)}{E(J-H)}$ and $\frac{E(J-H)}{E(J-K)}$. For the rest, the wavelength represents$\frac{E(\lambda-K)}{E(J-K)}$.
	\label{difexcol}}
\end{figure}

Using these values and those from Sections \ref{devexp}, \ref{exmir} and \ref{exbet}, we can check the residuals of each fit (Figure \ref{difexcol}) and the $\chi^2$ statistic (Table \ref{chiext}) to evaluate the behavior of both laws.
        
	\begin{deluxetable}{ccc}
		\tablecaption{$\chi^2$ for the color excess ratios derived from eqs. \ref{massa} and \ref{beteq}.
		\label{chiext}}
		\tablewidth{0pt}
		\tablehead{
			\colhead{Range} & \colhead{$\chi^2(\alpha$)} & \colhead{$\chi^2(\beta)$}
		}
		\startdata
			NIR &1.2&1.4\\
			MIR &23.8&25.1\\
			Total &24.9&29.1\\
		\enddata
	\end{deluxetable}
            
According to Figure \ref{difexcol} and table \ref{chiext}, it seems that both laws represent equally well the behavior of the extinction between 1.5 and 3 $\mu m$, while they underperform in the mid-infrared. Although this gives us a hint that there is a change in the extinction law beyond $3~\mu m$, there is also the chance that these differences arise from a poor calibration of the intrinsic colors of the RCGs. As we have shown previously, there is a systematic $0.05~mag$ error on the intrinsic colors that could add up to a $0.07$ error on the color excess ratios, swallowing completely any difference between laws in Figure \ref{difexcol}. This is unlikely to be the case: in the NIR this error is most likely to translate into a common offset (as metallicity will turn the stars redder or bluer) and in the MIR it will be greatly attenuated. Yet to completely roule out this hypothesis, we need a good calibration of the intrinsic colors of the RCGs at longer wavelengths.

Beyond calibration issues, the remaining possibility is a change in the extinction law beyond $3~\mu m$. This could translate into either a simple change in the slope of the law (as modeled by its exponent) or a change in the shape of the law (and hence the whole functional). To test this we have to take into account that following \citet{I05}, all the color ratios in the MIR are referred to $(J-K)$, effectively mixing both regimes. We can circumvent this by defining a new set of purely MIR color ratios, as can be seen in Table \ref{ratcolmironly}. We opt to use $([3]-[5])$ as a common denominator as it is the pair with the larger color excess for our sample, hence minimizing relative errors. We can see that the $([5]-[8])/([3]-[5])$ ratio yields a negative value. Yet it follows from Equation \ref{massa} that any color excess ratio:
\begin{equation}
E(\lambda_1-\lambda_2)\propto \frac{1}{1+(\lambda_{1}/\lambda_0)^{-\alpha}}-\frac{1}{1+(\lambda_{2}/\lambda_0)^{-\alpha}}
\end{equation}
it follows that since $\lambda_{[8]}>\lambda_{[5]}$ and $\lambda_{[5]}>\lambda_{[3]}$, the excess ratio is defined positive, no matter what choice of exponent. The reasoning holds also for a $\lambda^{-\beta}$ law. Although this effect is much less severe for all the other ratios, our selected laws are not able to reproduce these ratios. This matches what is shown in Figure 6 from \citet{I05}: lines of sight crossing regions of higher extinction deviate severely from the extinction law, particularly at $\sim 3~\mu m$, $\sim6~\mu m$ and $\lambda>8~\mu m$.

	\begin{table}[!h]
		\caption{Color excess ratios for GLIMPSE data.
		\label{ratcolmironly}}
		\begin{center}
		\begin{tabular}{cc}
			\tableline
			Ratio & Value \\
			\tableline
			([3]-[4])/([3]-[5])&0.23$\pm$0.08\\
			([3]-[8])/([3]-[5])&0.59$\pm$0.06\\
			([4]-[5])/([3]-[5])&0.71$\pm$0.06\\
			([4]-[8])/([3]-[5])&0.45$\pm$0.06\\
			([5]-[8])/([3]-[5])&-0.35$\pm$0.05\\
			\tableline
		\end{tabular}
		\end{center}
	\end{table}

If we restrict ourselves to the NIR, the model from \citet{FM09} yields a slightly better fit. Even if it requires one more free parameter (as the wavelength normalization $\lambda_0$ of Equation \ref{massa} expression needs to be tuned properly), as is noted in \cite{FM09}, the extinction from Equation \ref{massa} holds down to 0.3 $\mu m$, while the authors show that this is not the case for a $\lambda^{-\beta}$ law.
      
It should be noted that along these calculations no use of any reference wavelength (central, isophotal, etc.) for the filters is made, and instead the ratios are calculated through direct integration of the appropriate efficiencies, extinction laws and stellar synthetic spectra.

\subsection{$A_\lambda/A_V$ ratios}
    
Assuming a profile for the Johnson V filter, one can apply Equation \ref{refext} to derive the $A_{\lambda}/A_\mathrm{V}$ values displayed in Table \ref{extratallv}. While the color excess ratios do not depend on $A_\mathrm{V}$ or $R_\mathrm{V}$, $A_\mathrm{\lambda}/A_\mathrm{V}$ does so. This means that to get a realistic vale for this ratio we need to assume a meaningful value for them. As we do not have any a priori information about which $R_V$ or $A_\mathrm{V}$ best represent our data, we limit ourselves to evaluate $A_\mathrm{\lambda}/A_\mathrm{V}$ over all the full variation range used in Section \ref{devexp} and we obtain means and deviations using the posterior distribution (as it gives us those areas in the parameter space that correspond to our measurements with more probability) as a weighting function for all the ($R_V$,$E(B-V)$) combinations. These results are detailed in table \ref{extratallv}. The differences between the two laws come, in part, from the fact that these values are calculated extrapolating to $\sim0.5\ \mu m$, and we have noted that a $\lambda^{-\beta}$ dependence of the extinction with the wavelength may not be representative of this regime. The $A_\lambda/A_V$ are calculated mostly to offer an easy comparison with other works as \citet{C89} or \citet{RL85} (Table \ref{extratbib}).

	\begin{table}[!h]
		\caption{Extinction ratios derived using the law from \cite{FM09} ($\alpha$) and a power law ($\beta$).
		\label{extratallv}}
		\begin{center}
		\begin{tabular}{ccc}
			\multicolumn{3}{c}{$\alpha$}\\
			\tableline
			$Ratio$&2MASS&Sys. error\\
			\tableline
			$A_J/A_V$&0.197$\pm$0.009&0.02\\
			$A_H/A_V$&0.097$\pm$0.006&0.016\\
			$A_K/A_V$&0.048$\pm$0.003&0.011\\
			\tableline
			$A_H/A_J$&0.494$\pm$0.004&0.02\\
			$A_K/A_J$&0.243$\pm$0.003&0.02\\
			
			\noalign{\smallskip}
			\multicolumn{3}{c}{$\beta$}\\
			\tableline
			$Ratio$&2MASS&Sys. error\\
			\tableline
			$A_J/A_V$&0.167$\pm$0.025&0.012\\
			$A_H/A_V$&0.084$\pm$0.013&0.009\\
			$A_K/A_V$&0.040$\pm$0.007&0.006\\
			\tableline
			$A_H/A_J$&0.5046$\pm$0.0002&0.018\\
			$A_K/A_J$&0.2376$\pm$0.0016&0.019\\
			\tableline 
		\end{tabular}
		\end{center}
	\end{table}

	\begin{deluxetable}{ccc}
		\tablecaption{Extinction ratios for Rieke \& Lebofsky(1985) and Cardelli et al. (1989).
		\label{extratbib}}
		\tablewidth{0pt}
		\tablehead{
		\colhead{
			$A_{\lambda}/A_V$} & \colhead{RL85} & \colhead{C89}
		}
		\startdata
			J& 0.282&0.282\\
			H& 0.175&0.190\\
			K& 0.112&0.114\\
		\enddata 
	\end{deluxetable}

The two laws under consideration here come closer to agreement if we calculate the ratios against $A_\mathrm{J}$ (pointing to the effects induced when extrapolating to the visible wavelengths, which also may account for part of the difference with the classical coefficients).
          
In any case, our results, both $A_\mathrm{K}/A_V$ and $A_\mathrm{K}/A_\mathrm{J}$, differ significantly with those of the classic literature (Table \ref{extratbib}). Although some of the discrepancies are due to the differences in the photometric systems under consideration, it is obvious that these ratios are not representative of the extinction in the inner Milky Way. Although the main conclusions from studies such as \citet{CL08} or \citet{BG05} are not affected by this extinction ratio, these authors rely on $A_\mathrm{J}/A_\mathrm{K}$ to derive distances along the line of sight. Taking into account the results from this paper, these values should then be revised.
      
Values such as those of Table \ref{extratallv} are not unheard of. A growing number of papers point reach similar conclusions as those here exposed. \citet{G09} derive a ratio $A_\mathrm{V}/A_\mathrm{K}$ of $28.7\pm14$, which yields $\beta=2.64\pm0.52$, in good agreement with our values, although the authors support a model in which $\beta$ varies spatially over (l,b), behavior for which no statistical evidence is found here. We use 0.25$^\circ\times0.25^\circ$ fields, so our method is biased against small scale variations, like those found by the authors; it is possible, then, that these variations exist, but in a scale small enough to be smeared out in our data. \citet{NI06} found that $\beta=1.99\pm0.01$ for the bulge, which implies that $A_\mathrm{K}/A_\mathrm{J}=0.331\pm0.004$, close to our $\sim0.24$ value. The same authors, basing their work on simultaneous observations in V and the NIR, find the ratios $A_\mathrm{V}:A_\mathrm{J}:A_\mathrm{H}:A_\mathrm{Ks}=1:0.188:0.108:0.062$ \citep{NI08}; these values--directly estimated and not extrapolated--are closer to those derived using the \cite{FM09} model.
	
More recently, \citet{SH09}, using UKIDSS data for fields with $l>27^{\circ}$ infer $\beta=2.14\pm0.05$, an exponent that implies $R^{JH}_{HK}=1.86\pm0.11$, quite similar to our ratio of $R^{JH}_{HK}=1.79\pm0.09$.
      
Again, part of the variation between results will come from the inhomogeneous photometric systems, but the reader should take into account two aspects of the analysis. According to Equation \ref{refext}, some variation is expected to be caused by the different stellar sample used, as color excess ratios are dependent on the stellar population under study. These differences are, nonetheless, very small, and most of the time fall within the error bars. 

Also, and more important, most of the literature reviewed relies on the manipulation of Equation \ref{beteq} to obtain an expression that links wavelength, $\beta$ and ratios of the form $A_{1}/A_{2}$ or $R^{12}_{23}$. This relation is then fitted to a given set of data measured along one or various lines of sight. To do this, it is necessary to assume a representative wavelength for each passband against which ratios are calculated. The usual choices are the effective passband wavelength or the isophotal $\lambda$, calculated with some spectral distribution (this even has the added difficulty that the isophotal wavelength itself depends on the shape of the extinction and its absolute value, $A_V$). In the case at hand, the selection of ($\lambda_\mathrm{J},\lambda_\mathrm{H},\lambda_\mathrm{K}$) should be considered not only when calculating $\beta$, but also when comparing different results.
      
\begin{figure*}[ht!]
	\centering
	\includegraphics[width=6.cm,angle=90]{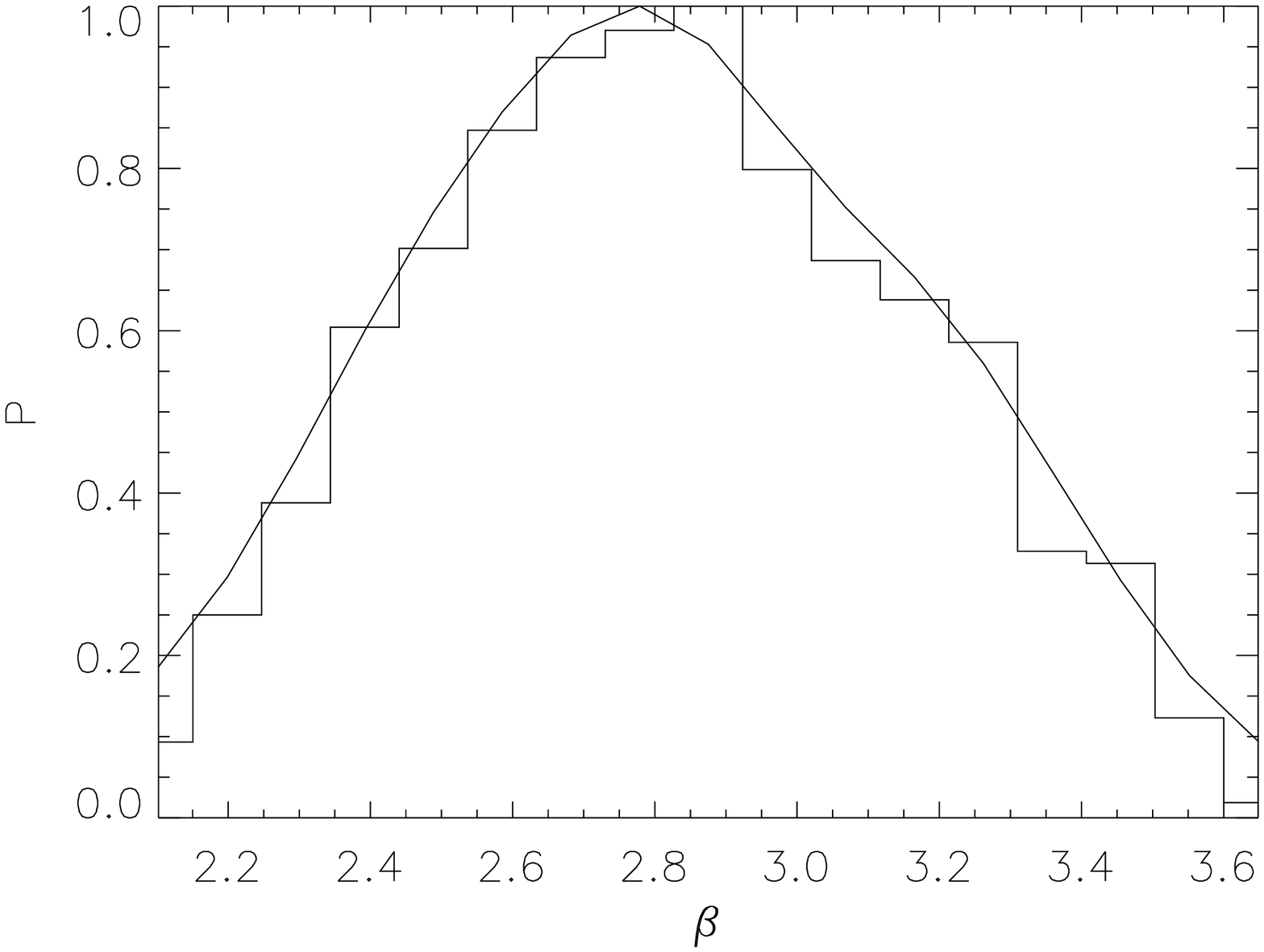}
	\includegraphics[width=6.cm,angle=90]{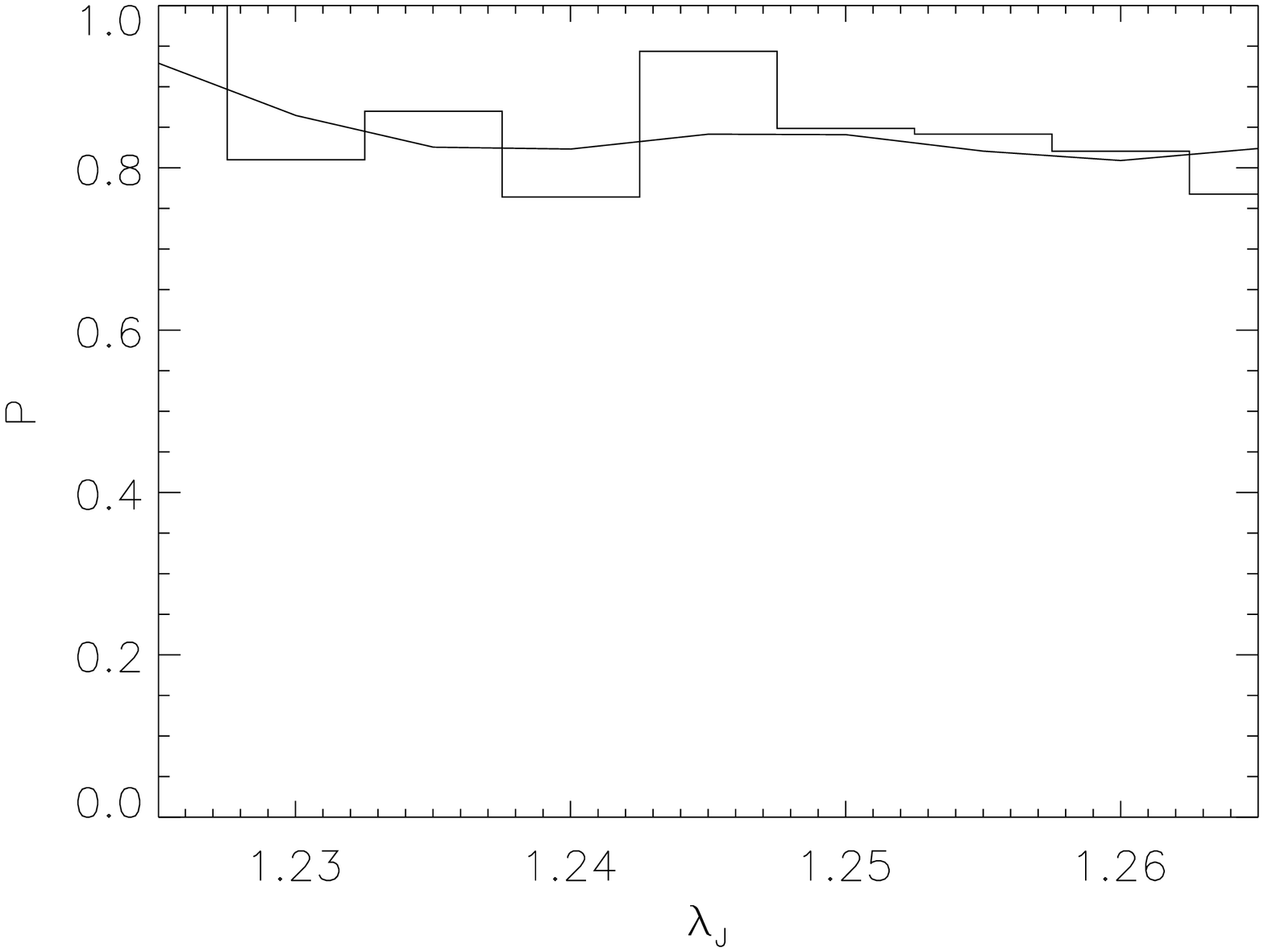}
	\includegraphics[width=6.cm,angle=90]{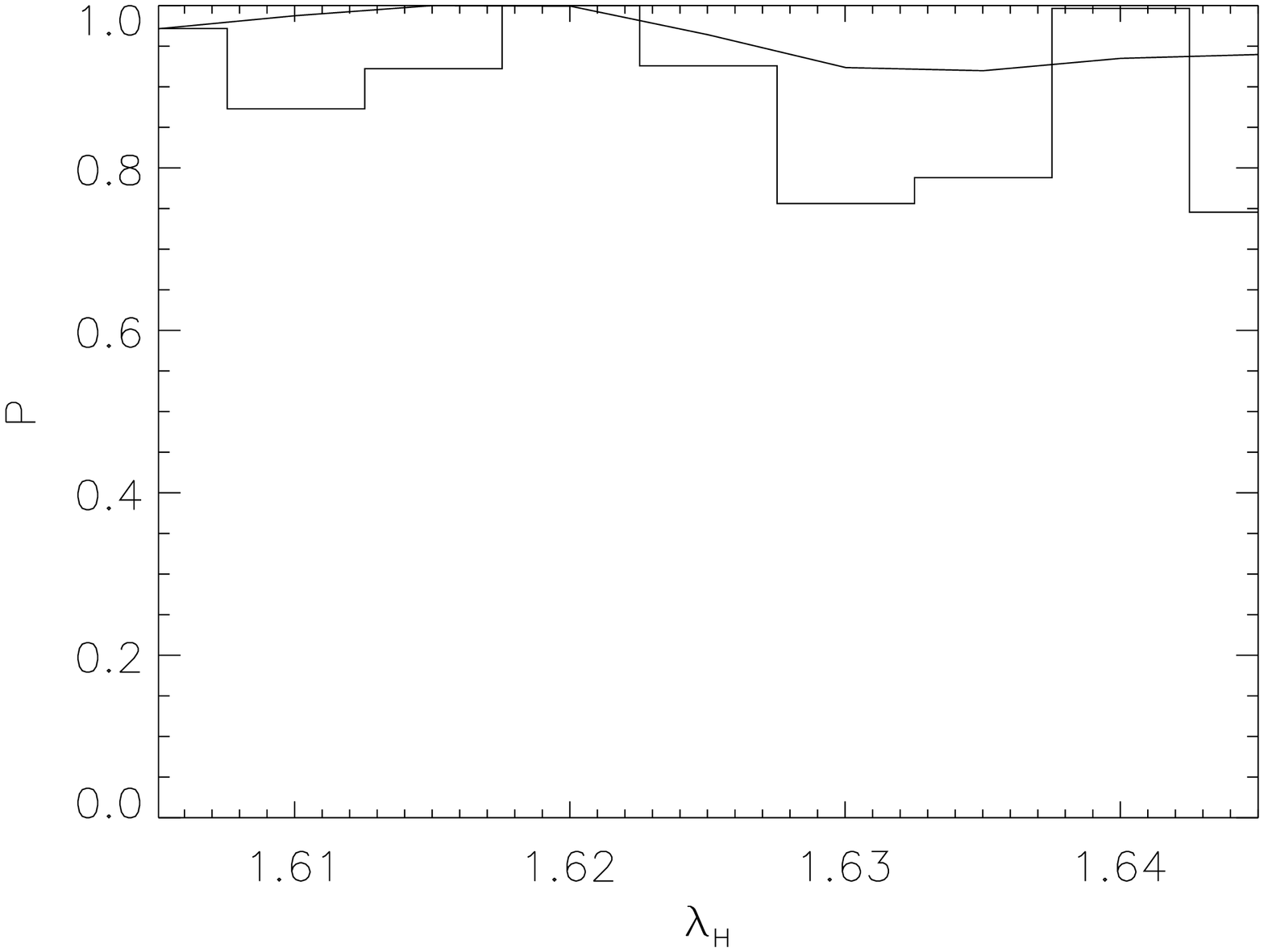}
	\includegraphics[width=6.cm,angle=90]{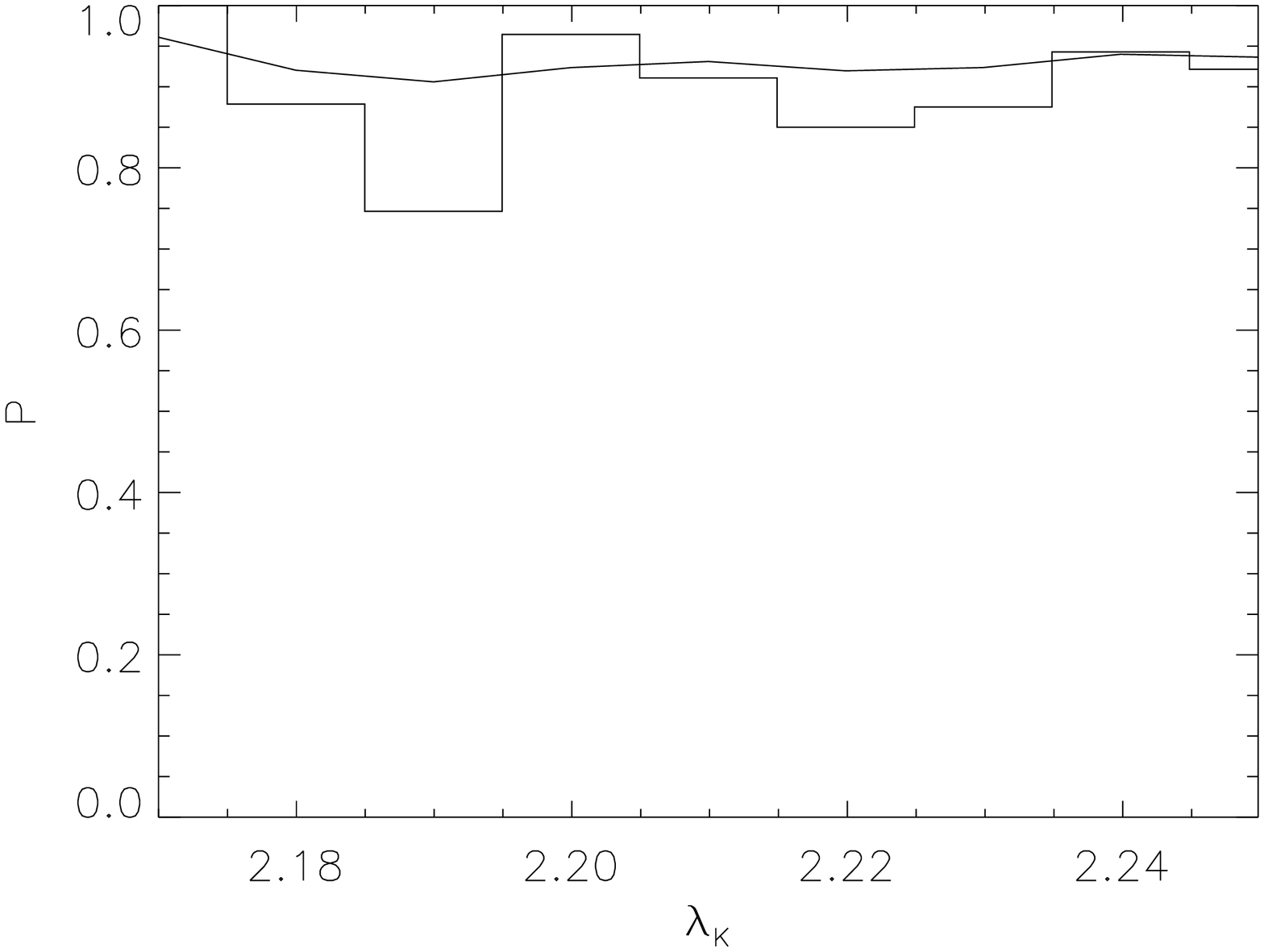}
	\protect\caption[ ]{Marginalized posterior distribution, normalized to its maximum, for the three wavelengths and the exponent $\beta$ of Equation \ref{ratbet}. For the implications of the shapes of these distributions, see text.
	\label{postebet}}
\end{figure*}

In fact, with Equation \ref{beteq} we can obtain the following relation:
\begin{equation}
\label{ratbet}
	R^{12}_{23}=\frac{A_{1}-A_{2}}{A_{2}-A_{3}}=
	\frac{\lambda_1^{-\beta}-\lambda_2^{-\beta}}{\lambda_2^{-\beta}-\lambda_3^{-\beta}}
\end{equation}
And with this and the results from table \ref{rats} for the near-infrared, one can set the three wavelengths as free parameters and do inference over them and $\beta$, in the very same fashion as in Section \ref{devexp}. This yields the results in Figure \ref{postebet}. While the marginalized posterior for the exponent is reasonably similar to a Gaussian, with $\mu=2.8$ and $\sigma=0.3$, the distributions for each wavelength are more or less flat. This implies that these three parameters are not tied by ($R^\mathrm{JH}_\mathrm{HK},R^\mathrm{JH}_\mathrm{JK},R^\mathrm{HK}_\mathrm{JK}$) and so for any combination of filter wavelengths  there is always an exponent within an interval spanning at least 0.8 units that can provide a good fit of the data. When marginalizing over the wavelengths, (i.e. taking the mean of all the solutions for any possible combination of $\lambda_\mathrm{J},\lambda_\mathrm{H}$ and $\lambda_\mathrm{K}$), the result comes in concordance with those in Section \ref{devexp}.
      
It follows that when comparing exponents obtained with different sets of wavelengths, say isophotal (as in \citet{I05}) or the central value for each filter \citep{SH09}, much of the observed variation can be due to the inhomogeneity of these choices.
      
The method used here evaluates the pertinent integrals of Equation \ref{refext} instead of relying in a fixed reference value of $\lambda$, and so it does not suffer of these problems.
      
\subsection{Probing deeper into the Galaxy with UKIDSS}

In these inner galactic fields, the depth of 2MASS is greatly diminished by the presence of high stellar density. Because UKIDSS uses a detector with higher spatial resolution, it is less affected by crowding, reaching deeper into the Galaxy. This survey is calibrated with 2MASS data, but as 2MASS uses a slightly different filter set, a magnitude transformation is needed between both photometric systems. Following \citet{hod09}, these transformations are:
\begin{eqnarray}
\label{schetrans}
	J_{UK}=J_{2M}-0.065(J-H)_{2M}+0.015E(B-V)\\
	H_{UK}=H_{2M}+0.07(J-H)_{2M}+0.005E(B-V)-0.03\\
	\label{schetrans2}
	K_{UK}=K_{2M}+0.01(J-K)_{2M}+0.005E(B-V)
\end{eqnarray}
The color terms over $(J-H)_{2M}$ and $(J-K)_{2M}$ are determined in low extinction fields, and the dependency on $E(B-V)$ \citep[derived from the maps of][]{schemap} tries to model the effect of interstellar extinction over the different band-passes of both surveys.

Overall, the calibration is good down to a few hundredths of a magnitude, but in fields such as the ones we study here, where the interstellar contribution is high, these equations introduce an extinction dependent error \citep[as][overestimates $E(B-V)$]{schemap} that will affect any derived extinction law. In fact, studies such as \citet{SH09} avoid the innermost lines of sight.

\begin{figure*}[ht!]
	\centering
	\includegraphics[width=6.cm,angle=90]{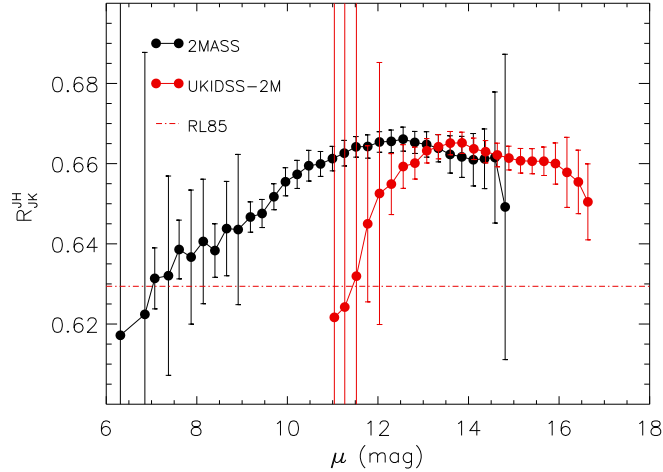}
	\caption{Variation of $R^\mathrm{JH}_\mathrm{JK}$ with $m_K$ averaged over all the lines of sight. Black dots are for 2MASS data and red dots for UKIDSS transformed into 2MASS system. The red line marks the ratio predicted by \citet{RL85}.
	\label{ratukidss}}
\end{figure*}

To directly compare UKIDSS ratios with 2MASS data, we need to translate magnitudes from the former into the later. If we want to avoid the outlined problem, this requires a star-by-star independent estimate of $E(B-V)$. It is possible to do so (or at least obtain a good approximation) using populations of know intrinsic color, such as RCGs, but it is a complex and very time consuming procedure. We can still use UKIDSS data, nonetheless, using our previously selected RCGs and obtaining with them a spatially averaged $E(J-K_\mathrm{S})$ with which we can transform extinction ratios from one system into the other. The results of this are presented in Figure \ref{ratukidss}. This procedure has the added bonus that it removes the extinction related calibration problems addressed in UKIDSS DR8\footnote{See http://surveys.roe.ac.uk/wsa/knownIssues.html}. As can be seen, once transformed into 2MASS system, UKIDSS ratios follow smoothly the behavior present in Figure \ref{ratall}, and the color excess ratio seems to remain constant (except for saturation/completeness effects). In fact, if we repeat the statistical analysis of section \ref{statana}, the results show that all UKIDSS ratios can be grouped into a single cluster. This means that the data can be described by a single mean value and there is no statistically significant variation.

\section{Conclusions}
    
In this work we study the large scale behavior of the infrared extinction on the inner Milky Way. We present a method to obtain extinction ratios that does not rely on the choice of wavelength for a given set of filters, and applying it to 2MASS and GLIMPSE data, we find that:

\begin{itemize}
	\item[-]The interstellar matter seems to be more transparent in the infrared than previously thought, and the traditional visible to infrared ratios of \citet{RL85} and \citet{C89} do not provide a good representation of its evolution over wavelength.
	\item[-]Using this method, there is no evidence for an azimuthal variation of the extinction law toward the innermost Galaxy ($0^\circ<l<30^\circ$). These variations are present either on a spatial scale that is too small (hence smeared out in our data) or too large (when comparing the inner and outer disk, for example).
	\item[-]This change happens in $R_{GC}$, as near the solar system the color excess ratios are quite similar to those of \cite{RL85}, while towards the Galactic center these change to higher values. Although a physical explanation for this is beyond the scope of this paper, an obvious hypothesis is to tie this change in behavior with the first cross of a spiral arm, where high density molecular regions are likely to affect light in a different way than local, diffuse interstellar medium.
	\item[-]Neither a single exponent for a power law nor the expression from \cite{FM09} seem to be able to account for the overall shape of $A_{\lambda}$ between 1 and 8$\mu m$, although a proper calibration of the intrinsic colors for the  red clump stars is needed to further assess this claim. 
\end{itemize}

All of this has profound implications on what we know about the structure of our Galaxy, particularly regarding its large-scale geometry, as much of the studies on the bulge and bar assume a much smaller $A_{J}/A_{K}$ ratio than the one found here. This implies that the distances estimated using this ratio will be underestimated if compared when compared to the values that could be derived with the $A_{J}/A_{K}$ from this paper. It can be easily shown that:
\begin{equation}
\frac{d}{d_{\mathrm{RL}}}\sim10^{0.34\cdot E(J-K_\mathrm{S})}
\end{equation}
Where $d_{\mathrm{RL}}$ is the distance derived with the extinction law from \cite{RL85} and $d$ the same value but using the results from this paper. For very reddened fields, this ratio can be as high as 2. Yet there is an extra element that we should take into account: as can be seen in Figure \ref{ratall}, there is evidence that the color excess ratio (and also $A_{J}/A_{K}$) is not constant along the line of sight and that for the first few kiloparsecs from the Sun, values from \cite{RL85} work well. Being so, we cannot do a straight comparison between distances along the line of sight derived using the classical extinction law and the one presented here, as there is only a fraction of the light path where they differ.
      
\acknowledgments

	Financial support by the Spanish Ministerio de Ciencia e Innovaci\'on (MCINN) under AYA2010-21697-C05-5, AYA2008-06166-C03-3 and by the Spanish Ministry of Economy and Competitiveness through project AYA2010--18029 (Solar Magnetism and Astrophysical Spectropolarimetry) is gratefully acknowledged. AAR also acknowledges financial support through the Ram\'on y Cajal fellowship. Partially funded by the Spanish MICINN under the Consolider-Ingenio 2010 Program grant CSD2006-00070: First Science with the GTC (http://www.iac.es/consolider-ingenio-gtc).

	This publication makes use of data products from the Two Micron All Sky Survey, which is a joint project of the University of Massachusetts and the Infrared Processing and Analysis Center/California Institute of Technology, funded by the National Aeronautics and Space Administration and the National Science Foundation. 

	The UKIDSS project is defined in Lawrence et al (2007). UKIDSS uses the UKIRT Wide Field Camera (WFCAM; Casali et al, 2007). The photometric system is described in Hewett et al (2006), and the calibration is described in Hodgkin et al. (2009). The pipeline processing and science archive are described in Irwin et al (2009, in prep) and Hambly et al (2008).

	This work is based in part on observations made with the Spitzer Space Telescope, which is operated by the Jet Propulsion Laboratory, California Institute of Technology under a contract with NASA.






\clearpage






\end{document}